\documentstyle[aasms4,12pt]{article}
\raggedright
\newcommand{\elec}{$\rm e^{-}$}

\newcommand{\mo}{$\rm \mu _{0}$}

\newcommand{\Mbar}{$\bar{M_I}$ }
\newcommand{\mbar}{$\bar{m_I}$ }

\newcommand{\mss}{mag arcsec$^{-2}$}

\newcommand{\plm}{$\pm$ }

\newcommand{\lta}{$\leq $}

\newcommand{\Bmu}{\rm $\mu_B \:$}

\newcommand{\Bmo}{\rm ${\mu _B}$(0) }

\newcommand{\alp}{$\alpha$\ }
\newcommand{\etal}{{\it et.al.}\ }

\newcommand{\app}{$\sim$}

\newcommand{\Msol}{$M_{\odot}$\ }

\newcommand{\lsb}{\rm low surface brightness}

\input{epsf}
\input{rotate}
\begin{document}
\baselineskip=14pt
\def\oneskip{\baselineskip\baselineskip}
\pagestyle{myheadings}
\parskip=12pt
\def\oneskip{\baselineskip\baselineskip}
\floatsep 0.5in
\textfloatsep 0.5in

\title{ HST WFPC2 Imaging of Three Low Surface Brightness 
Dwarf Elliptical Galaxies in the Virgo Cluster}
\author{Karen O'Neil\altaffilmark{1}}
\affil{Arecibo Observatory, HC3 Box 53995, Arecibo, PR 00612}
\affil{\it koneil@naic.edu}
\altaffiltext{1}{Work was done while at the University of Oregon}
\author{G.D. Bothun}
\affil{Dept. of Physics, University of Oregon, Eugene OR, 97403}
\affil{\it nuts@bigmoo.uoregon.edu}
\and
\author{C. D. Impey}
\affil{Steward Observatory, University of Arizona, Tucson AZ, 85721}
\affil{\it impey@as.arizona.edu}

\begin{abstract}

\noindent Utilizing the F814W and F300W filters, Hubble Space Telescope
Wide Field Planetary Camera-2 (WFPC2) images were taken of three
low surface brightness dwarf elliptical galaxies in the
Virgo cluster.  The intent of the observations was to determine
the small scale structure in these enigmatic galaxies, and to attempt
to learn something about the nature of their giant branch through
the detection of luminosity fluctuations.  In two of the three studied galaxies, V7L3 and V1L4, 
the luminosity fluctuations
in the inner, constant surface brightness regions were unambiguously detected.
At the nominal distance of the Virgo cluster, the measured luminosity
fluctuations in the F814W band yields a density of 2 -- 10 red giants/pixel.
In the most extreme of these two cases, V7L3, we derive a surface density of giant
stars of $\sim$ 3 per 10 pc$^2$.   Using the observed $B-V$ and $V-I$
colors as a constraint, we could find no model that would reproduce the
observed fluctuation signal and blue colors if there was a significant
population of M-giants in these systems.  
Overall, our results are consistent
with a mean spectral type of K0 -- K2 which implies a relatively metal
poor population.  The third system, V2L8, did not have a detectable fluctuation
signal which possibly implies it is not in the Virgo cluster.  
Interestingly, this system is highly nucleated.  Our observations have
resolved this nucleus and if V2L8 is in Virgo, then
we have discovered what is likely the smallest bulge measured to date,
having an effective radius of only 50 pc.  This bulge
is quite red (as red as giant ellipticals) and its entirely possible
that this nucleated dE galaxy, in fact, is a very large galaxy located
in the background.  As such, it is highly reminiscent of the manner 
in which Malin-1 was discovered.  Optical spectroscopy of this nucleus
is required to confirm this.  Finally, we find no evidence for small scale
clumping of stars in any of the studied systems at this much improved spatial resolution.
This implies these systems are dynamically well-relaxed and that the
physical cause of their observed low surface brightnesses is their
low density.  When
imaged at the high spatial resolution of the WFPC2  ($\sim$ 6 pc per pixel),
the galaxies are easy to look right through without evening knowing
they are present in the very middle of the WFPC2 frame.  They appear only
as elevated ``sky noise''.

\end{abstract}
\baselineskip=24pt
\section{Introduction}
Little is known or understood about the current stellar populations and/or
star formation histories of low surface brightness (LSB) dwarf elliptical
(dE) galaxies.  What we do know from various studies (e.g. 
Sung \etal 1998; 
Jerjen \& Dressler 1997; Secker 1996; Durrell \etal 1996; Meylan \&
Prugniel 1994; Lee, Freedman, \& Madore 1993; Peterson \& Caldwell
1993; Impey, Bothun \& Malin 1988; Caldwell 1987; Caldwell \& Bothun 1987;
Bothun \etal 1986; Kormendy 1985, 1987; 
Bothun \etal 1985; Bothun \& Caldwell 1984;)
can be summarized  as follows 
(see Ferguson \& Bingelli 1994 for a fuller review):

1) LSB dEs in the Virgo and Fornax clusters generally define a tight 
surface brightness-magnitude relation (Secker \& Harris 1996).  
This relation is driven 
by the tendency for dE surface brightness profiles to be extremely
well-fit by an exponential function, coupled with a near constancy of
the disk scale length ($\alpha$ $\sim$ 0.9 $\pm$ 0.1 kpc) (see Bothun,
Caldwell \& Schombert 1989; Caldwell \& Bothun 1987; Young and Currie
1995).  Thus
variations in luminosity are driven solely by variations in  central
surface brightness.   In a simple universe there would also be a 
corresponding color vs. surface brightness relation, with the lower
luminosity dEs being redder than the higher luminosity ones.  This
would make the  surface brightness-magnitude relation merely a 
fading sequence.  Naturally, things are more complicated than this
as no color-central surface brightness relation has been observed
for any sample.  In fact, the available data actually define a relation
in the opposite sense, namely dEs with the lowest central surface
brightness are the bluest (see Figure 3 in Bothun, Impey \& Malin 1991).  

2) There is a small but important component of very LSB dEs with
large scale lengths that strongly deviate from the standard surface
brightness-magnitude relation (see Impey, Bothun \& Malin 1988;
Bothun, Impey, \& Malin 1991; Caldwell, \etal 1998; O'Neil 1997).  Though there is no difference in
mean color, these very diffuse dEs may be fundamentally
different than the other dEs.  Some of the more extreme examples in this class reach
central surface brightnesses as low as \Bmo = 26.0 \mss\ but have scale
lengths of \app 1.5 kpc. 

3) An appreciable fraction of LSB dEs have conspicuous nuclei.
Spectroscopy (e.g. Bothun \& Mould 1988; Brodie \& Huchra 1991;
Peterson \& Caldwell 1993; Held \& Mould 1994)
indicates a stellar population similar to that of metal-rich galactic
globulars but with stronger Balmer line equivalent widths, perhaps
indicating a lower mean age.  In general there is little difference
in color between most nuclei and the surrounding envelope.  Whether
these nuclei are mini-bulges (e.g. r$^{1/4}$ components) or the site
of a secondary star formation event is currently unclear.

4)  Most dEs have little neutral hydrogen, suggesting that  substantial
gas loss may have occurred as the result of baryonic blowout in shallow potentials due to
energy input from supernovae (e.g. Dekel \& Silk 1986; Vader 1987;
Silk, Wyse \& Shields 1987; Spaans \& Norman 1997).

5) While some possible ``transition'' objects have been identified on their
way to becoming gas poor dEs (e.g. Meurer, Mackie \& Carignan 1994; Knezek,
Sembach \& Gallager 1997; Vader \& Chaboyer 1994; Sage, \etal 1992; 
Conselice \& Gallagher 1998) its fairly unclear 
what their present evolutionary nature is.  Only a handful of these candidate
transition galaxies exist, compared to the
relatively large numbers of dEs in clusters, suggesting that whatever
evolutionary process has produced dEs is no longer ongoing with much frequency.

6)  The number density of dEs in groups and clusters seems to be correlated
with the total cluster luminosity in the sense that large brighter clusters (e.g. Virgo) have
significantly more dEs than 
fainter clusters such as Fornax (see Ferguson 1991; Secker \& Harris 1996).  
This is a compelling result which strongly suggests some quite macroscopic 
physical event is responsible for the production of dEs in clusters.
Indeed, very deep studies of the Coma cluster suggest that there may be
thousands of dEs in that environment (Ulmer \etal 1996; Bernstein \etal 1995;
Secker, Harris, \& Plummer 1997)

Missing from the above list is any explanation as to why the surface brightnesses of these dEs
can be so low, at a wide range of $B-V$ and $V-I$ colors.  
Surface brightness is, of course, a convolution of the average
separation between the stars and the luminosity function of the stars in
the galaxy.  Broadly speaking, the available photometric data on dEs is 
inconsistent with a significant change in the stellar luminosity function --
that is, the broad-band integrated colors as well as
nuclear spectra indicate the light is dominated
by a giant branch augmented by A,F and G main sequence
stars.  Given this, the most probable reason that these galaxies have such
low surface brightnesses is a larger than average separation
between the stars, or between individual red giants in the case of giant
dominated integrated light.  Is this a formation effect?  That is, have these
systems always been of low mass density (see de Blok \& McGaugh 1997) or
has there been some profound evolutionary process, perhaps associated with
significant mass loss (e.g. Dekel \& Silk 1986) that has ``puffed'' what 
once were compact galaxies into a considerably more diffuse state?

But what is the evidence, apart from general broad band colors, that 
the light from dE galaxies is giant dominated.  As remarked by Bothun,
Impey, \& Malin (1991) and McGaugh \etal (1995), there is particular difficulty
in fitting stellar population models to the blue end of the LSB dE
sequence, because these objects are blue in the clear absence of on-going
star formation.  In general, this end is defined by objects with $B-V$ $\sim$
0.4 $-$ 0.5 and $V-I$  $\sim$ 0.6 $-$ 0.8.  For these objects, their
colors can be reproduced using a large population of A,F and G stars,
and a reduced giant branch (which indicates a young mean age for the
galaxy), thus bringing into question the statement that their light
is giant dominated.   
One way to directly test whether these blue dEs still have
giant dominated light is offered by the measurement
of luminosity fluctuations using the Hubble
Space Telescope (HST).   Previous
attempts to measure the fluctuation signal of LSB dEs from the ground have
been successful.  Bothun \etal (1991) successfully detected the B-band
fluctuations in two LSB dE galaxies in Fornax using a detector with
pixel size of 0.33 arcseconds under conditions of 0.7-0.8 arcsecond
seeing.  Jerjen \etal (1998) measured the R-band fluctuation signal 
for a few dE galaxies in Sculptor using a detector with pixel size of
0.60 arcseconds under conditions of 1.5 arcsecond seeing.  

Clearly, at 0.1 arcsecond per pixel and a PSF of approximately 0.2
arcseconds, HST observations using the Wide Field Planetary Camera-2 
(WFPC2) present a unique opportunity for a robust measurement
of the fluctuation signal from LSB dEs in structures as distant as
the Virgo cluster.  A priori, what one might expect from such
measurements? Well suppose there is some dE with a region of constant
surface brightness of B = 25.0 \mss\ a few arcseconds in size.
At I, the mean surface brightness will be approximately I = 23.5 \mss.
At 0.1 arcsecs per pixel, each WFPC2
pixel would have B = 28.5 mag.  At the distance of  the Virgo cluster
(m$-$M = 31.5 for this illustrative purpose only), the absolute magnitude per 
pixel in the I-band is $-$3.0.
If the light per pixel is giant dominated then  this absolute magnitude level
is reached with just 2--10 giants, depending on their spectral type.   
The Poisson noise associated with
such a discrete distribution of giants would indeed be large ($\geq$ 33\%).
If, on the other hand, the light from these blue dEs is dominated by
F and G main sequence stars, then several hundred per pixel are required
and the corresponding fluctuation signal would be significantly reduced.
We thus seek to determine the amplitude of the fluctuation signal in 
a small sample of blue LSB dEs in Virgo to 
a) directly test that the light from these dEs is still giant dominated and 
b) to show membership in the Virgo cluster.

Additional motivation
for performing these observations is three-fold: 1) There is 
conflicting information in the literature concerning the metal abundance
and/or effective temperatures of the giant branches in these systems.   
For instance, the interpretation offered by Bothun \& Mould (1988) is 
somewhat different than that put forth by Brodie \& Huchra (1991).
By measuring the fluctuation signal, we have an opportunity to infer 
the approximate K-to-M ratio in the composite giant branch.
In globular clusters, it has been shown (Reed, Hesser \& Shawl 1988), that
the K/M giant ratio is a good indicator of metallicity.
2) We know
very little about the small scale structure of these enigmatic dwarf galaxies.
For instance, are these dEs of low surface brightness because the mean
giant luminosity per pixel is low or is the actual surface density of
giants (absolute numbers of stars per pixel) low?  Determining the
luminosity fluctuations associated with discrete numbers of giants per
pixel can help resolve this.  3) The nature of the nuclei which frequent
many dEs in Virgo remains unclear.  One dE in our sample exhibits a
very red nucleus that is spatially unresolved from the ground.  WFPC2
observations may help to resolve this nucleus to better determine its
nature.

In this paper we describe our imaging experiment of three LSB dEs in Virgo. 
This experiment has never been tried before on Virgo dEs.  
Section 2 describes our dE sample as well as the instrumentation 
and data reduction procedures.  In section 3 we 
report on the detection of the fluctuation signal and present some
model analysis on the nature of the composite giant branch in
these systems.
Section 4 gives a complete error analysis for the images and
section 5 discusses the nature of the individual dEs in more detail.

\section{Observations and Data Reduction}

\subsection{The dE Sample - Global Properties}

For this study we selected three LSB dEs in Virgo from the ground-based
Las Campa\~nas 2.5m Dupont telescope CCD sample of Impey, Bothun, \& Malin
(1988; IBM hereafter).  Two of these three are in the Virgo Cluster 
Catalog (VCC) of Sandage and Binggeli (1984).   The objects chosen are
V1L4 (VCC1582), V2L8, and V7L3 (VCC1149).  

V1L4 is fairly easy to identify in the 
ground based images, but the presence of a large number of (apparent) foreground
stars makes analysis of this galaxy, at  ground-based resolution, difficult.
The extrapolated central surface
brightness is 24.2 B \mss, and the integrated B magnitude is
16.7, making it the brightest of the three dEs in this study.  The
galaxy is circular in appearance, with the hint of a faint spiral arm
on the north-eastern side of the galaxy (much like the incipient spiral
structure of Malin 1 -- see Impey \& Bothun 1989).
The surface brightness profile consists
of a flat central region followed by an exponential fall-off.
This lack of a true exponential profile, to R=0,
prevents an accurate determination of scale length but its
27.0 B \mss\ isophote diameter suggests a scale length
similar to the other two dEs in this study.

V2L8 has a  central surface brightness of only 25.8 B \mss\ making it
the most diffuse object in this study (and explaining why it is not
a VCC object).  The galaxy is roughly circular in appearance
but is well nucleated.  Unfortunately, in the IBM data 
there is a CCD flaw running through the center of the galaxy 
which prevented much analysis of this nucleation.  We have included this
object in our sample in hopes of resolving the nucleation with HST.
V2L8 nominally has a scale length slightly larger than the typical dE 
in Virgo or Fornax ($\alpha_{V2L8}$ = 1.2 kpc).  When combined with the
very low central surface brightness , V2L8 is well outside
the standard surface brightness-magnitude relation discussed in Section 1.

V7L3 is intermediate between the other two.  It has a measured 
central surface brightness of 25.1 \mss\ and a scale length of
\alp$_{V7L3}$ = 1.1 kpc.  Like V2L8, it too is an exception to
the standard surface brightness -- magnitude relation of dE galaxies.
As will be seen, V7L3 was the most difficult galaxy to identify in the 
WFPC2 data because its very diffuse and lacks any nucleation.

The light distribution of the three galaxies is remarkably similar.  Once the 
bright nuclear core of V2L8 is removed, all three dEs have a flat 
inner surface brightness profile followed by an exponential profile which 
continues through the detection limit.  It is in these flat diffuse
regions that we  seek to measure the fluctuation signal.  These regions
may be kept diffuse by the action of
background radiation pressure, stellar winds, 
or some other mechanism that provides enough outward pressure to prevent
an increase in density and reduction in scale size of these diffuse 
regions (i.e. Kepner, Babul \& Spergel 1997).

Information from the ground based images are given in Table 1, 
and described below.
All quantities were calculated using the Johnson B band filter unless otherwise noted.
It should be stated that although the images and zeropoints used for this table
are the same as those used in IBM, the parameters have been independently calculated, by re-doing the surface photometry.
\begin{description}
\item {\bf Columns 1 and 2:} Galaxy names as given in IBM (Column 1) and in the Binggeli,
Sandage, and Tarenghi atlas (1985) (Column 2).
\item {\bf Columns 3 and 4:} RA and Dec of the galaxies, as found using the STSDAS 
METRIC task on the WFPC2 F814W images (J2000 epoch).
\item {\bf Column 5:} Central surface brightness, in \mss.
\item {\bf Column 6:} The scale length, in arcsecs, as defined in equation~\ref{eq:mue} (below).
\item {\bf Column 7:} The total B magnitude integrated out to the 27.0 \mss\ isophote.
\item {\bf Column 8:} The isophotal diameter measured at the \Bmu = 27.0 \mss\ level.
\item {\bf Columns 9 and 10:} The B $-$ V and V $-$ I colors, measured through the d=20''
aperture for V2L8 and V1L4, and through the d=34'' for V7L3, due to the difficulty in 
obtaining an accurate color at smaller apertures (see section 5).  The errors are
0.05 and 0.1 for B $-$ V and V $-$ I, respectively.
\end{description}

The colors of these dEs are fairly blue.  For comparison, the 
typical Galactic globular cluster has colors of B $-$ V = 0.62 \plm 0.02 and 
V $-$ I = 0.93 \plm 0.05 ([Fe/H] \lta  $-$1.7).   It is likely that the stellar
populations in these dEs are metal-poor with younger mean age than those
found in galactic globulars.  This would imply a deficit of
M-giants which is something that can be constrained from the measured
fluctuation signal.

\subsection{Instrumentation}

The WFPC2 consists of three Wide Field cameras and one Planetary camera.  The Wide Field
cameras have a focal ratio of f/12.9 and a field of
view of 80" x 80" with each pixel sub-tending 0.0996 arcsec$^2$.  The three cameras form an L-shape,
with the Planetary camera completing the square.  The Planetary camera
has a focal ratio of f/28.3, 
0.0455 arcsec$^2$/pixel, and an overall field of view of 36 arcsec$^2$.  All four cameras have
an 800 x 800 pixel silicon CCD  with a thermo-electric cooler to suppress dark current.
The WFPC2 has two readout formats -- single pixel resolution (FULL mode) and 2x2 pixel binning
(AREA mode).
The digital to analog converter used a gain of 7 e$^-$/digital number.  

The data for this survey was acquired 
on 1 May 1996, 3 August 1996, and 3 October 1996.  Each field was chosen
so that the center of the dE was located in the WF3 image.
Four images of each galaxy were taken
using all four WF and PC chips, for a total of 2100s and 2200s through the F300W and F814W 
filters, respectively.  
The F814W filter is a broadband filter with $\lambda_0$ = 7924 \AA\  and $\Delta \lambda_{1/2}$
= 1497 \AA.   It is designed to be similar to the Cousins I-band filter.  The F300W 
filter has $\lambda_0$ = 2941 \AA\ and $\Delta \lambda_{1/2}$ = 757 \AA, and is the WFPC-2
wide band U filter.  The
F814W images were taken in FULL mode, while the F300W images were taken in AREA mode.
Because of the CCD response, the S/N through the F814W filter was considerably higher than through
the F300W filter.  Surface brightness profiles and  structural parameters
were all found through the F814W images. 
Figure~\ref{fig:wfpc} shows the full (mosaicked) images through the
F814W filter.

Sky flat fields of the sunlight Earth were taken through each filter and routinely calibrated
against an internal flat field calibration system.  The internal system
consists of two lamps (optical and UV) illuminating a diffuser plate.
The internal flats are used to monitor and correct for changes in the
flat fields.  Dark fields are averages of ten calibration frames taken over the space
of two weeks.  The intrinsic dark rate of the WFPC2 CCDs is \lta 0.01 e$^-$/pixel/sec.
A bias field was generated for each image using 
extended register pixels which do not view the sky. 

The data reduction process was as follows: First, all known bad pixels were removed, using
the static mask reference file.  
The bias level was then removed 
from each frame.  The bias image, generated to remove any position-dependent bias pattern,
was then subtracted from the image, as was the dark field image.  Flat
field multiplication was then performed.
All the above image calibration was performed at STScI using the standard WFPC2-specific 
calibration algorithms (the pipeline).  
After the images were reduced,
they were inspected for obvious flaws such as filter ghosts or reflections.  As none were
found, all the images were used in the subsequent analysis.
Each frame was then shifted, registered and combined, using the STSDAS CRREJ
procedure to eliminate cosmic rays and
other small scale flaws.  The resultant 2100s -- 2200s
images were then checked by eye to insure any registration
errors were less than 0.5 pixel.

\subsection{Data Reduction}

The zeropoints for each field were taken from the PHOTFLAM value given in the image headers.
The zeropoint, in the STMAG system (the space telescope system based on a spectrum with
constant flux per unit wavelength set to approximate the Johnson system at V), is 
\[\rm ZP_{STMAG}\:=\:-2.5 log(PHOTFLAM)\:-\:21.1.\]
For the F814W filter, the PHOTFLAM was 2.5451 x 10$^{-18}$, corresponding to a zeropoint of
22.886.  For the F300W filter the PHOTFLAM was 
6.0240 x 10$^{-17}$, with a zeropoint of 19.450.  Conversion 
to the Cousins I band was done using the value given in by Whitmore in the
{\it WFPC2 Photometry Cookbook} of I $-$ F814W = 1.22 \plm 0.01 (for objects
with the colors of galaxies).
Conversion from the F300W band to the Johnson U band is more
complicated due to an imperfect match between the filters.  As a result, 
we used the value obtained by O'Neil, Bothun, \& Impey (1998) of U $-$ F300W = 0.04 \plm 0.1.

The physical center of each galaxy, estimated by centroiding with respect to outer isophotes,
was found and ellipses were fit around that point to obtain the
intensity in each annulus using the modified GASP software (Cawson 1983; Bothun \etal 1986).
The pixel size of the survey provides a seeing radius (stellar psf) 
of 0.1'' for the Planetary camera,
and 0.2'' for the Wide Field camera.  The average 
sky-subtracted intensity within each (annular) ellipse was found and calibrated with the
photometric zeropoint.  Background galaxies were masked with the GASP software, which
sets the value of the affected pixel to -32768 and subsequently ignores the affected region.

Exponential surface brightness profiles were plotted against the major axis (in arcsec) for
each galaxy, using the following equation: 
\begin{equation}\rm \Sigma (r)\:=\: \Sigma_0\:e^{-r \over \alpha} \label{eq:sige}\end{equation}
where $\Sigma_0$ is the central surface brightness of the disk in linear units
(\Msol/pc$^2$), and \alp is the exponential scale length in arcsec.  This can also be written
(the form used for data analysis) as
\begin{equation}\rm \mu (r)\:=\:\mu (0)\:+\:({1.086 \over \alpha})r  \label{eq:mue}\end{equation}
where \mo\ is the central surface brightness in \mss.  

The average sky brightness through the F814W filter was 23.01 \mss\ (which
corresponds to about 21.8 \mss\ in the Johnson I-band system).
An accurate (error \lta  0.25 \mss) radial surface brightness profile 
was typically found out to 25.5 \mss\ (10\% of the sky background).

\section{Data Analysis and Modeling}

\subsection{Measuring the Fluctuation Signal}

The flat surface brightness profile in the inner core of these dE galaxies, 
combined with the exceptionally flat sky background of our WFPC2 F814W images 
(flat to less than 0.1\%), 
allows for an accurate  detection of luminosity fluctuations caused by
the stellar population of these inner regions.
Figure~\ref{fig:core} shows the grey-scale images
for the inner regions of
the galaxies in the F814W images. In all three cases the
profiles are flat, with mean $\mu$ = 24.39, 26.23, and 25.49 F814W \mss\ for 
V1L4, V2L8, and V7L3, respectively.
Pixel-to-pixel variations within the flat regions (as defined in 
Table 2, Column 2), as well as for the sky,
were then found by determining the mean electron count and dispersion in 
three sets of 135 boxes 5, 10, and 15 pixels wide, for a total of 47,250 pixels,
which were spread randomly throughout the 
region of constant surface brightness (see also Bothun, Impey, \& Malin 1991). 
Multiple random samplings of these regions were done so that errors 
could be determined via statistical bootstrap techniques.

The intrinsic fluctuation signal was found by subtracting, in 
quadrature, the r.m.s. variation of the sky (still in e$^-$) from that 
within the the constant surface brightness regions.    It is precisely
the existence of regions of constant surface brightness that encompass
several thousand pixels that allows for the fluctuation signal to be
measured in such a straight forward manner.  That is, the fluctuation
signal can be extracted without any need to Fourier analyze
the image to recover the power spectrum, as is traditionally done
in studies such as these.  As will be shown below, this technique has
allows the fluctuation signal to be measured to high accuracy when
it is detected in this manner. 

For these observations,
the sky background averaged 145--170 electrons which is well above
the readout noise for WFPC2.  In the absence of other sources of
noise (e.g. filter fluorescence, CTE problems, scattered light -- see
section 4 for details)
the only other contribution to the fluctuation signal besides the galaxy 
comes from the Poisson noise in the sky background.   
Division by the average intensity of
the constant surface brightness region then gives the fractional
luminosity fluctuation which is presumably driven by a Poisson distribution of
red giant stars per pixel.
However, there is one small complication which make this whole procedure
a bit less than straight forward and that is the simple fact that the
angular extent of the galaxy (at very faint isophotes) 
is comparable to the WFPC2 field of view.
Examining the outer isophotes from the Las Campa\~nas I band image
reveals that, at the maximum radii available for the WFPC2 images
(r=80'' for V2L8, V1L4 and r = 100''
for V7L3), the annular surface brightness is
26.85, 28.83, and 27.69 \mss\ for V2L8, V1L4, and V7L3, respectively.
This implies that only the light from V1L4 has fallen off enough to render it 
insignificant (e.g. $<$0.5\%) in the calculations of both the sky brightness and 
its r.m.s. variation.  For the two other galaxies,
2.5\%, and 1.3\% of the measured sky value is a contribution from the outer
stellar light in V2L8 and V7L3, respectively and needs to be accounted
for in the determination of the true sky value. 

In the case of sky limited exposures,
such as we have, the r.m.s. sky error in electrons is:
$${\sqrt{(sky\:intensity)\:+\:(number\:of\:exposures)*(read\:noise)^2}\:=
\:\sigma_{rms}\:(sky,\:in\:e^-).}$$
If we assume that this r.m.s. error represents the true
sky noise for all three galaxies (that is, the sky error found is the true
$\sigma_{rms}\:(sky)$), we can determine the true galaxy r.m.s. error using
$${{\sqrt{\sigma_{measured}(galaxy)^2\:-\:\sigma_{rms}(sky)^2}}\over
{\left[ galaxy\:intensity\right] \:-\:\left[ sky\:intensity\right] }}\:=\:\sigma_{true}(galaxy)\: .$$
Uncertainties are dominated by the uncertainty in the numerator.
A statistical bootstrap method is used to determine the uncertainty
in the measured values of $\sigma_{measured}(galaxy)$ and
$\sigma_{rms}(sky)$.   These values can be found in Table 2 (see below).
In general $\sigma_{rms}(sky)$ is larger than the r.m.s
of the actual sky counts, in electrons, indicating that readout noise
is still a component in the overall noise profile of both the galaxy
and sky images.  

After grinding through this procedure for all three galaxies, we measure
the fractional luminosity fluctuations to be 0.42 \plm 0.53,
0.33 \plm 0.05, and 0.65 \plm 0.17 for V2L8,  V1L4, and V7L3, respectively.
The fluctuation signal for V2L8 clearly is not statistically significant, but
to first order the large and statistically significant fluctuation signal 
measured for V1L4 and V7L3 confirms would what was introduced
in Section 1 (see also Figure 5 in Bothun, Impey, \& Malin 1991).

Combining this measure of the luminosity fluctuations with 
the probable distance modulus to
Virgo yields an estimate for the average magnitude of the stars  producing the
observed fluctuation (see Tonry \& Schneider 1988).  Of course,
the distance modulus to Virgo is uncertain and values of m$-$M = 31.0 --
31.5 remain consistent with the data (see Bothun 1998).
Using this range of distance moduli,
we can determine the absolute magnitude/pixel for the constant surface brightness 
regions.  For example, in our 2200 second combined exposure,
V1L4 has counts of 181.2 \plm 0.8 \elec/pixel versus 
145.2 \plm 0.3 \elec/pixel for the
sky or a net count of 36 \plm 0.9 \elec/pixel which converts to a
mean magnitude/pixel of 28.26 in the Cousins I band
($m\: =\: -2.5log\left[{181.2-145.2}\over{7}\right]\:+\:31.26\:-\:1.22\:=\:28.26$).  The measured
fluctuation signal of 0.33 implies that, on average, there are
10 giants per pixel.
Using m $-$ M = 31.0 then, gives \Mbar = $-$0.24.   For V7L3 we
derive a mean magnitude/pixel of 30.01 with an average of 3
giants per pixel.  This yields \Mbar = $-$0.55.  These values 
are significantly below the typical values of \Mbar found for
luminous ellipticals (see below).

\subsection{Modeling the Giant Branch}

We now have enough information to approximately model the giant branch
in terms of a mixture of giants of spectral type K and M, together with
an underlying main sequence of A, F, and G stars.  One way to determine
our model is to simply appeal to the calculations of Worthey (1994) in
which the fluctuation magnitude is listed for a variety of stellar
populations of differing ages and metal abundances.  However, those models were
developed for application to giant ellipticals and its not clear if
they are appropriate for our dE galaxies for the following statistical
reason: In a giant elliptical at the distance of the Virgo cluster,
each pixel would contain several hundred giants (and a total of several thousand
stars) and thus each pixel represents a statistically
reliable realization of the general stellar population.  In our case,
this is simply not true as each pixel contains a very small number
of giants (certainly less than 10 and maybe as low as 2) and hence
we are subject to discrete effects.  In the extreme, part of our
fluctuation signal may in fact be driven by the tendency for some
pixels to contain zero giants.  Thus, we are in a much different counting
regime than the case of a giant elliptical.

Nonetheless, we begin with an inspection of the Worthey models.
In the I-band, \Mbar decreases with increasing metallicity
for a fixed age population.   Its only at
near-IR wavelengths that
the fluctuation magnitude starts to rapidly increase as you get to
more metal rich populations which contain the cooler, luminous
M-giants.   In addition, throughout the regime of
low metallicity ($-$2.00 $\leq$ [Fe/H] $\leq$ 0.0), \Mbar is relatively
constant.  In this metallicity regime, \Mbar $\sim$ $-$1.8 $\pm$ 0.1 over
the age range 8--12 Gyr.   This is well above the values we found
from our data, which are at most \Mbar $\sim$ $-$1.0 for m$-$M = 31.5.
So with respect to our data, these models are extremely poor fits in
that they achieve  \Mbar $\sim$ $-$0.5 only in metal rich cases but those
populations have $V - I$ $\sim$ 1.3 -- 1.4.  Conversely, using the bluest
$V - I$ models ($V-I$ = 0.86 corresponding to [Fe/H] = $-$2.00 and
age of 8 Gyr) yield  \Mbar $\sim$ $-$1.95.  So
the comprehensive models of Worthey 
do not appear to have any applicability to dE galaxies if 
\Mbar is mostly driven by metallicity variations; we simply cannot
even come close to getting consistent values for both $V - I$ and \Mbar.

To make further progress we model the giant branch by adopting
the following procedure:  1) 
The measured luminosity fluctuation to first order fixes the number
of giants per pixel; 2) We
assume the giant branch can be populated by stars of spectral type
K0 thru M2; 3) We adopt absolute magnitudes and colors for giants
as a function of spectral type as shown in Table 3 (not considering
types later than M2 as they are typically found in metal rich bulges,
a state far removed from the dEs.); 4) We use the
observed B $-$ V and V $-$ I colors as additional constraints which
help us to evaluate the contribution of A0 -- F0 stars to the
integrated light.

For a specific demonstration of this procedure we take the case of
V1L4 at assumed (m$-$M) = 31.0.  The observed fluctuation signal of 33\% argues 
for 10 giants per pixel, to first order.  This yields \Mbar = $-$0.24 which
is approximately the same as for a K0 giant.  Since the observed color of V1L4
is bluer than that of a K0 giant in  V $-$ I, then there clearly is
an important contribution from an underlying bluer population.  Hence,
we seek an approximate model for the giant branch and the ratio of
giant branch to AFG stars that can simultaneously 
satisfy the color and fluctuation signal constraints, within the
observed errors.   These AFG stars represent a blue underlying population
which could be a populated main sequence or a blue horizontal
branch population.

As an example of an acceptable fit, a model (Model A) with 2 A0 and 30 F0 main
sequence stars in combination with 2 K0, 2 K2 and 2 K3 giants returns
\Mbar = $-$0.30, $B-V$ = 0.56, $V-I$ = 0.81 and m$-$M = 31.26.   Another model (Model B)
with 3 A0 and 40 F0 stars in combination with 4 K3 giants returns
\Mbar = $-$0.34, $B-V$ = 0.46, $V-I$ = 0.75 and m$-$M = 31.31. 
Both of these models return distance moduli estimates consistent with
cluster membership.  In section 6 we will apply the $V-I$ vs \Mbar
calibration of Tonry (1991) and Tonry \etal (1997) to uncover 
widely inconsistent results strongly suggesting that, like the Worthey
models, the calibration for giant ellipticals does not hold for
these galaxies.

Clearly, given the
accuracy of the measurements we can only come up with only approximate
models, but the particular feature
we are interested in constraining from these observations is the mean
spectral type (effective temperature) of the giant branch.  
By
gauging this we will have another handle on the metallicity of the
stars in these systems.  The combination of the observed fluctuation
signal and the color does have high constraining power in this regard.

As a further example, we can take Model A above and add a 5\% (by number)
contribution of M2 stars.  This yields
\Mbar = $-$0.97, $B-V$ = 0.59, $V-I$ = 0.92 and m$-$M = 31.43.  This
is not very consistent with the data and in particular \Mbar is too
bright and $V-I$ is marginally too red.  To reduce \Mbar while still
retaining M2 giants, requires the addition of A0 and F0 stars.  This
addition will make the broad band colors bluer but will also increase
the distance modulus as the absolute magnitude per pixel is now
increased.  If we double the contribution  of F-stars
we obtain
\Mbar = $-$0.79, $B-V$ = 0.50, $V-I$ = 0.83 and m$-$M = 31.67.   Thus,
we can only accommodate a small M-star contribution in V1L4 for the
largest probably distance modulus to Virgo.  For the shorter distance modulus
no M-giant contribution can be accommodated.  Furthermore, none of our models
actually can get as blue as $V-I$ = 0.7 while being consistent with
the derived \Mbar (see Table 2).  For instance, using G5 giants can drive  $V-I$ down
to 0.7 but such models consistently return values for \Mbar that are fainter
then we observe (see also Worthey 1994).

\subsection{Overall Results}

Table 2 lists our overall results in terms of determining \Mbar and its
error.  All values relevant to the calculation of the
fluctuation signal are given in units of electrons per pixel.

The table is laid out as follows:

\begin{itemize}
\item{Column 1:} The galaxy name.

\item{Column 2:} The radius range over which the flat surface brightness profileholds.

\item{Column 3:} The average central surface brightness, through the F814W filter, for the studied regions.

\item{Column 4:} The average central surface brightness, converted to the I band (Section 2.3).

\item{Column 5:} The average galaxy+sky counts within the region defined by Column 2, in electrons.

\item{Column 6:} The r.m.s. error ($\sigma$) for Column 5.

\item{Column 7:} The average sky counts for each image, also in electrons.

\item{Column 8:} The r.m.s. error ($\sigma$) for Column 7.

\item{Column 9:} The luminosity fluctuation, from electron counts, determined 
for each galaxy, followed by an error estimate (detailed in section 4).

\item{Column 10:} The absolute fluctuation magnitude (\Mbar) 

\end{itemize}

The results of these calculations summarized in this table are clear.
The high resolution and low noise of the WFPC2 has allowed for a
reliable determination of the luminosity fluctuation signal in 
2 out of 3 cases.  The amplitude of this signal is large for the cases of
of V1L4 and V7L3 and are likely 
produced by only 2--10 giant stars/pixel depending
on the types of giants considered.

In Table 4 we list the best fitting stellar population models to
the observed color and fluctuation data.  These models were obtained
by averaging the results of all models that gave values of
$B-V$, $V-I$ and \Mbar that were within the errors in the data and
which produced a distance modulus in the range (m-M) = 31.0 $-$ 31.7. 
No model that we ran 
got as blue as $V - I$ = 0.7.  Table 4 is laid out as follows

\begin{itemize}
\item{Column 1:} Galaxy name

\item{Column 2:} Mean spectral type of Giant Branch

\item{Column 3:} K/M giant number ratio if allowed by the data

\item{Column 4:} A+F/K+M number ratio

\item{Column 5:} distance modulus

\item{Column 6:} $B-V$

\item{Column 7:} $V-I$

\item{Column 8:} \Mbar

\end{itemize}

\section{Error Analysis}

The variance, as measured in electrons, is typically 5--10\% higher in the
constant surface brightness regions of the dE galaxies compared to the
sky background.  This is the fluctuation signal but, before we can
directly associate that with a Poisson distribution of giant stars
per pixel, we must gain a thorough understanding of potential
systematic errors arising from the WFPC2 system.  These other potential
sources of error are:

\begin{itemize}
\item {\bf Dark glow}:  This is a non-uniform background which may appear on the WFPC-2 chips and
is due to luminescence in the MgF$_2$ CCD windows under cosmic ray bombardment.  Examination for
this effect can be done through looking for a small intensity curvature across the sky.  This
effect is not found in the WFPC-2 images discussed in this paper.

\item {\bf CTE errors:}  WFPC-2 chips experience a charge transfer efficiency (CTE)
loss across the chip of up to 20\% along the Y-axis. This effect, however, is readily
reduced by long exposures and high DN counts.  The combined images discussed in this paper
are the equivalent of 2200s images, providing raw (non-averaged) counts of
2,000 DN (14,000 e$^-$) in a 5x5 pixel box.  This reduces the effect of CTE errors
from the 20\% mark to 2\% -- 3\% (i.e. Whitmore 1998).  Additionally, the majority of the
CTE loss occurs at the edges of the chips, and can readily be seen in a plot 
of the average sky counts along a chip column.  When this was done on the data discussed herein,
it was determined that virtually all of the loss occurred in the 50 pixels at the edges of the chips.
These pixels were therefore eliminated from the analyzed image, further
reducing any CTE problems to under 0.5\%.

\item {\bf S/N loss at the chip edges:} Within approximately 50 pixels of the inner edges of the 
wide field chips the signal-to-noise ratio drops considerably due to vignetting and 
spherical aberration as the light is divided between two chips.  With the images
in question, this effect can be readily eliminated by again examining the sky counts in the
chip's inner regions.  Eliminating the inner 50 pixels from each image reduced the effects
of this problem to zero.

\item {\bf Geometric Distortion:}  Geometric distortion near the edges of the chips result in
a change in the surface area covered by each pixel.  In general, this effect
is not relevant for surface photometry where azimuthal averages are
taken and the variance in the sky background is determined over areas
encompassing thousands of pixels.  
The flat fields also reduce this problem considerably
by boosting the values of the smaller pixels.  
By analyzing a large number of sky/galaxy regions, each containing
a minimum of 25 pixels, we again reduced this effect to under 0.1\%.

\item {\bf Scattered Light:} Bright stars whose light falls on the planetary camera pyramid mirror
can produce an obvious artifact on the CCDs, typically in the shape of a large arc.  None
of the images in this paper suffered from this effect.

\item {\bf Systematic Errors:} Potential systematic errors could arise from inappropriate
box sizes resulting in  under or over-sampling, 
the underlying galaxy surface brightness not being constant,
and the presence of point sources in the studied region.  To counter the first problem, the
variance was computed for three difference sized boxes (5x5, 10x10, and 15x15 
pixels) and the results 
compared.   The weighted differences in sky and galaxy counts between the three box sizes was
under 0.2\%.  
To look into the possibility that 
the studied galaxy regions may not have been flat, a comparison can be done between the 
counts found in the inner and outer portions of the studied regions.  In this case the errors remain under 0.5\%.

\end{itemize}

The cumulative result of these other effects
per statistics box results in a potential additional
photometric error of up to 1 -- 2\%.  However, our sky and
galaxy fluctuation signals and
their errors are determined by averaging over approximately 50,000
pixels (in 135 individual boxes) across the WF3 chip and hence
these additional errors are ultimately reduced to well under 1\%.  
The difference in luminosity fluctuation between the sky and
the galaxy signals ($\sim$ 5--10\%) is 
well above the level of any possible systematics.

\section{The Individual Galaxies}

The distribution of the giant stars in the inner regions of all three
LSB dEs appears to be completely uniform 
(Figure~\ref{fig:lum}). There are no
apparent clumps or clusters seen at our physical resolution scale
of approximately 15 pc.  Interestingly, because of the
low number of giant stars per pixel, WFPC2 imaging essentially renders 
these galaxies transparent and their presence appears only as a
``sky fluctuation'' (see also Figure 1).   Unless WFPC2 observers are
careful, they may well have an object like this in their field without
even knowing it.

\subsection{V1L4}

The ground-based data showed a number of bright regions or clumps in
this object.   However, it is clear from the higher angular resolution
WFPC2 data that these regions are mostly
background galaxies shining through V1L4 and hence the underlying
structure of V1L4 is quite smooth.
The background galaxies are described more fully
in another paper (O'Neil, Bothun, \& Impey 1999) and demonstrate
the transparent nature of this and other dE galaxies.  A
few other ``knots'' on the arc-second scale
can be identified which could be localized 
regions of star-formation.  Confirmation of this, however, can not be
provided by the F300W filter observations as that data is
extremely noisy.

Analysis of the luminosity fluctuations, described in the last 
section, show the typical star within V1L4's nuclear region to have 
\Mbar = $-$0.32 -- $-$0.82, (m$-$M = 31.0 -- 31.5) which corresponds to 
spectral types K0 through K2 in the mean.  Our models, however, do
accommodate the possibility a small M-giant contribution to the fluctuation
signal of K/M = 30, provided m$-$M = 31.5. 
If the luminosity distance is lower,
K/M goes to $\infty$, that is, the possibility of any M-type stars existing within
this galaxy goes to zero.  Keeping the results consistent with the observed V $-$ I color
does not change these results, which equate to 13 \plm 1 giant stars in a
10 pc$^2$ region of the galaxy, of which at most 0.5 could be an M giant star.
Figure~\ref{fig:lum}(a) shows the core of V1L4, with the region of flat surface brightness
lying in the defined annulus.  The contour lines in black demark the regions whose
brightness is at least 1 $\sigma$ above the mean surface
brightness in that region,
and thus probably are reflective of the actual
distribution of the individual giant stars.  Interestingly, rather than
being evenly distributed throughout the annulus, the
majority of the giant stars in this region appear to lie in the southern part of
V1L4's core, accounting for V1L4's slightly off-center appearance when
imaged at coarser angular resolution.
Additionally, it should be noted that the higher intensity regions do appear to
be grouped, indicating perhaps an old stellar cluster now traced by
the remnant giant population.

\subsection{V2L8}

Figures~\ref{fig:lum}(b) and (c) show the inner
regions of V2L8, with the region of constant surface brightness again 
de-marked by white circles and regions 1 $\sigma$ above the galaxy brightness 
defined by black contour lines.  The distribution of giant type stars
appears fairly even throughout the galaxy.
Analysis of the luminosity fluctuations of V2L8 outside the nucleated
region does 
not provide a statistically significant result, 
with $\sigma_{galaxy}$ = 0.42 \plm 0.53.  One possible reason for this
is that galaxy completely fills the WFPC2 field of view and no sky
measurement is possible.  Indeed, inspection of Table 2 shows that
the sky counts are significantly higher for this object, although
observing conditions (e.g. variable shuttle-glow, sun angle) could
also be responsible for these increased counts.  Given the strong
detection of the fluctuation signal for the other two dEs in our
sample, perhaps this null result indicates that V2L8 is background.
That might help to explain why it does not conform to the surface-brightness
magnitude relation and why it has a nucleus.  Recall, that a previously
nucleated dE in Virgo turned out to be  Malin 1 (Bothun \etal 1987).

WFPC2 imaging has clearly resolved the core of V2L8 in the 
F814W images although the core drops out entirely in the F300W image.
IBM measured V $-$ I = 1.9 through a 5 arcsecond diameter aperture.    
We have reanalyzed the data in attempts to better remove the bad Column
and re-measure the nuclear colors,  
but the measurements are quite sensitive to choice
of center.  Overall we find colors consistent with the IBM value but can
better demonstrate the uncertainty.  Based on this we conclude that the
V $-$ I color of the nucleus is 1.85 $\pm$ 0.15 mag, which is well within
the range defined by luminous ellipticals.  Thus, the nucleus of this 
dE galaxy is extraordinarily red, although the envelope of the galaxy
appears fairly blue.  

But what is the nature of this conspicuous red
core?
Fitting an $r^{1/4}$ profile gives an effective
radius ($r_e$) of 0.7 arcseconds and and effective surface brightness
of 22.0 \mss.  
Its F814W magnitude, as measured
through an aperture of diameter 2 arcseconds is $\sim$ 22.8.  
If V2L8 is in the Virgo cluster,
then this nucleus is,
in fact, an extremely small scale bulge with $r_e$ $\sim$ 50 pc and has
an absolute magnitude at Cousins I of $-$10 to $-$10.5, consistent with
it being a bright, very metal-rich globular cluster (perhaps
similar to those seen in NGC 5128 -- Frogel 1984).  Given
the extremely diffuse nature of the central regions of this object,
the formation of a highly compact bulge is very curious.  If true,
this is
the first identified $r^{1/4}$ component of a dE galaxy with such a
small scale length. The
red color further suggests a metal-rich giant population.  Attempts
at spectroscopy of this nucleus in February 1998 using the now defunct
MT were unsuccessful due to weather and difficulty in finding the
nucleus on the acquisition TV.   The lack of an observed fluctuation
signal, however, has renewed our quest for optical spectroscopy as
this object may be background and, like Malin 1, intrinsically large.

\subsection{V7L3}

The WFPC2 images show V7L3 to have a very even stellar distribution, with
even its core hardly brighter than the sky background.  Remarkably, 
even with the WFPC2 image (additively) binned in 10 x 10 pixels
(giving the image a resolution of 1''/pixel),
V7L3 is still a fairly diffuse blob within the sky image and quite difficult to identify.  The observed
fluctuation signal is relatively large (owing to its lower surface
brightness compared to V1L4)
and is consistent, to first
order, with a stellar population of only 3 giants per pixel, yielding 
\Mbar = $-$0.56 to $-$1.06, or spectral type  K2/K3.
This is a slightly later spectral type
than the case of V1L4, even though both dEs have the same
$V -I$ color.  To accommodate this requires a large contribution, per
pixel, from the underlying A and F stars (mostly F-stars).  However,
it is clear that the data can not accommodate M-giants (which have
$M_I$ $\sim$ $-$2.4), as the model quickly gets too red.
Moreover, the absolute magnitude per pixel in
the center regions is fainter than the absolute I-band magnitude of
an M2 star, which would lead to fluctuations which are larger 
than we observe.  In fact, it is very difficult to fit any one of our
seven component models to the data for this galaxy at the short distance 
modulus.  Successful models tend to be absurd (a point noted earlier 
by Bothun \etal 1991 regarding the colors of some of these dEs) and
require approximately equal mixtures of A0 main sequence stars and 
K-giants.  For instance model C has equal numbers of A0 and K3 stars
(and nothing else) and this returns
\Mbar = -0.74, $B-V$ = 0.54, $V-I$ = 0.85 and m$-$M = 31.12. Adding
10 times as many F stars to this model produces
\Mbar = -0.88, $B-V$ = 0.45, $V-I$ = 0.82 and pushes m$-$M to 
31.70.  Once  again its essentially impossible to push these models
as blue as $V-I$ = 0.70 while simultaneously reproducing the observed
\Mbar.

Since the composite giant branches appear
similar, the more diffuse nature of V7L3, relative to V1L4 must be
due directly to a lower surface density of giants or, equivalently,
an increased average spatial separation between giants stars.  The
physical cause of this is unclear.
Figure~\ref{fig:lum}(d) shows the core of V7L3 with the regions 1$\sigma$ above the
galaxy brightness demarcated by black contour lines.
Figure~\ref{fig:lum}(d) shows V7L3 to have the most even distribution of giant
stars of the three galaxies in this study.  This even stellar distribution
within V7L3's core, combined with the circular appearance of the galaxy and
the lack of any large stellar knots within V7L3 argues for the idea that 
LSB galaxies are diffuse and \lsb\ by nature, and not due to outside influences
that might cause the galaxies to ``puff-up'' in some stochastic manner.  
Under that scenario, one might expect there to be considerably more
clumpiness in the stellar distribution than we actually observe,
which in all three cases is consistent with an old, dynamically
relaxed distribution of giant stars.

\section{Discussion} 

The primary result of this study is the firm detection of luminosity
fluctuations which are associated with a small number of giants per pixel
in two of the three LSB dE galaxies in our sample.   Specifically,
luminosity fluctuations of the inner, constant surface brightness regions, 
yields a density of 2-10 red giants/pixel for two of the imaged galaxies. 
Since the distance to Virgo is 
relatively well known, we can use the measured fluctuation signal,
in combination with the observed $V - I$ color, to constrain the 
respective contributions of K and M giants to the observed light.  
In so doing, the result is clear.  We can not simultaneously account
for the observed fluctuation signal and the very blue $V - I$  in
any model that has an M-giant contribution.  In fact, the models
strongly favor very early K-giants and hence a relatively warm effective
temperature for the composite giant branch.  This implies the population
is relatively metal poor.  

In more general terms we find that its extremely difficult for any model
to reach $B-V$ $\sim$ 0.5 with $V-I$ as blue as 0.7 
yet still exhibit \Mbar brighter than $-$0.3.
This is relatively easy to understand as to achieve
such blue colors requires the addition of many F-stars (main sequence
or blue horizontal branch stars) which greatly
increases the number of stars per pixel and lowers the overall fluctuation 
signal.  So, in this sense, the stellar populations of these blue
LSB dE galaxies remain mysterious and ill-constrained.  
This has been noted as far back
as Bothun and Caldwell (1984) and is a manifestation of the basic
dilemma involved in trying to produce galaxies with $B-V$ $\sim$ 0.5
that have no active star formation and very low surface brightness.
The most confident statement we can make, from the fluctuation data,
is that the giant branch is likely devoid of a significant population of 
M-stars.

We can, of course, turn the situation around and derive the
distance to the Virgo cluster.  Two calibrations are available 
for this purpose.  Tonry (1991) gives
\[M_I(Cousins)\:=\: -4.84\: +\: 3.0(V\:-\:I)\]
based on a sample that includes colors as blue as V-I = 0.85.
The revision of this calibration, by Tonry \etal (1997), based on including
very red galaxies (and strictly valid only over the range 1.00
$\leq$ $V-I$ $\leq$ 1.30) is
\[M_I(Cousins)\: =\: -1.74\: +\: 4.5*[(V\:-\:I)\: -\: 1.15]\]
For V1L4  we derive \mbar = 30.76 and for V2L3 we get
\mbar = 30.45.  Both dEs have $V -I$ = 0.7 $\pm$ 0.1.
The Tonry (1991) calibration thus yields m$-$M =  33.3 $\pm$ 0.3
for the two galaxies averaged.  The Tonry \etal (1997) calibration
results in a distance modulus one magnitude farther.  If we believed
these calibrations, then these objects are clearly not in the Virgo
cluster.  However, this is more likely indicating that the metallicity
driven variation in \Mbar, which is at the heart of the calibration (and
the Worthey models), just does not apply to LSB dEs possibly due to
discrete effects.  At some level, the actual surface brightness (e.g. the number
of stars per pixel) becomes important.  Consider the extreme case where
either the surface brightness is sufficiently low, or the pixel size is
sufficiently small, that, on average, there is only 1 giant per pixel.
So the surface density of giants is now 1 $\pm$ 1 and the fluctuation
signal would be 100\%.  In this limit, it is not clear that the \Mbar 
vs $V-I$ calibration means anything because the dominant driver
of the fluctuation signal is the fact that some pixels would have zero
giants in them.    The case of V7L3, where we derive a surface density of giant
stars of $\sim$ 3 per 10 pc$^2$, is close to this limit.

Of course, in this limit, the color fluctuations on the pixel scale
would also be very severe.  We had hoped to measure this effect with
the combination of the F300W and F814W filters but were effectively
thwarted by the low S/N in the F300W case.   Without this additional
information, our constraint on the stellar population per pixel is limited
and all we can really do is focus on the relative contributions of 
K vs M giants.  In general, we find that we can not simultaneously 
produced the inferred pixel density of giants and the observed
$V-I$ color with any model that includes M-giants.   Another way to
state this is by again comparing our results with the models of Worthey (1994).
While  it is possible to match our observed spectral fluctuations 
with Worthey's predictions,  our galaxies still remain significantly
bluer in V $-$ I than the model predicts.
Since the redder colors of Worthey's models
are due in large part to the presence of late K and M giant stars, this offers
further evidence against a significant population of such stars
within V1L4 and  V7L3.  The apparent paucity of these stars
is likely an indication that these dE galaxies are relatively metal poor.    

For the case of V2L8, we did not detect a fluctuation signal.  While
this may be due to its large angular extent on the WF3 frame, it might
also indicate that V2L8 is background to Virgo.  If indeed V2L8 is
in the Virgo cluster then
we have discovered what is likely the smallest bulge measured to date,
having an effective radius of only 50 pc.  This bulge
is quite red (as red as giant ellipticals) and thus may well be
substantially more metal-rich than the rest of the galaxy.  Possible,
it is a signature of a secondary star formation event that occurred
over a very small spatial scale.  To date, no other dE LSB galaxy
that has been studied shows such a very small, very red core.
Clearly, spectroscopy of this core is desirable.  Either we have
a very small bulge here, or V2L8 is in the background and may therefore
by like Malin 1; a LSB object with an L*, metal-rich bulge (see
Impey and Bothun 1989).

Finally, we comment on the LSB nature of these objects.  
We find no evidence for small scale clumping of stars on the 10-20
parsec spatial scale. To first order, this
suggests these systems are dynamically relaxed.
Expansion of these systems is then unlikely to be the explanation for
their observed low surface brightnesses.   Since we have detected
surface brightness fluctuations coming from a very small number of
stars per pixel, then we know that individual giant stars are dominating
the light per pixel.  Thus, their LSB nature is also not caused by
an absence of giant light.  While this is not a surprising result, 
this study is the first to demonstrate that directly.  This leaves
the physical separation between individual giant stars as the cause
of the observed low surface brightnesses.
In the WFPC2 data, such low density
galaxies could easily be dismissed as ``sky noise''
and remain undetected.  The continuing difficultly to detect faint, LSB
galaxies with any instrumentation 
has clear implications for reliable
determinations of the faint end slope of the galaxy luminosity function.

We acknowledge HST award GO-05496 to help support data acquisition
and reduction.  We also acknowledge NSF support for low surface brightness
galaxy research at the University of Oregon.  Conversations with
Harry Ferguson on the general subject of luminosity fluctuations have
greatly improved our understand of the subject.

\clearpage
\centerline{\bf References}
Bernstein, G. M., Nichols, R. C., Tyson, J. A., Ulmer, M. P., \& Whittman, D.  1995 AJ 110, 1507\\

Binggeli, B., Sandage, A., \& Tarenghi, M 1985, AJ, 90, 1681\\

Bothun, G.D. 1998 {\it Modern Cosmological Observations and Problems}, Chapter 2\\

Bothun, G.D., Impey, C.D., \& Malin, David F.  1991 ApJ 376, 4\\

Bothun, G.D., Caldwell, N., \& Schombert, J.  1989 AJ 98, 1542\\

Bothun, G.D., \& Mould, J.R. 1988 ApJ 324, 123\\

Bothun, G.D., Mould, J.R., Caldwell, N., \& MacGillivray, H.T.  1986, AJ, 94, 23\\

Bothun, G.D. Mould, Jeremy R., Caldwell, Nelson, \& MacGillivray, Harvey T.  1986, AJ, 92, 1007\\

Bothun, G. D., Mould, J. R., Wirth, A., \& Caldwell, N. 1985 AJ 90, 697\\

Bothun, G. D., \& Caldwell, C. N.  1984 ApJ 280, 528\\

Brodie, Jean P., \& Huchra, John P. 1991 ApJ 379, 157\\

Caldwell, Nelson, Armandroff, Taft E., Da Costa, G. S., \& Seitzer, Patrick 1998 AJ 115, 535\\

Caldwell,  C. N., 1987 AJ 94, 1116\\

Caldwell,  C. N., \& Bothun, G. D.  1987 AJ 94, 1126\\

Cawson, M. 1983, Ph.D. thesis, University of Cambridge\\

Conselice, C. \& Gallagher 1998, MNRAS 297L, 34\

De Blok, W. J. G., \& McGaugh, S. S. 1997 MNRAS 290, 533\\

Dekel, A., \& Silk, J. 1986 ApJ 303, 39\\

Durrell, P., \etal 1996 AJ 112, 972\\

Ferguson, Henry C., \& Binggeli, Bruno 1994 A\&ARv 6, 67\\

Ferguson, H.C.  1991 BAAS 23, 1338\\

Frogel, J. 1984 ApJ 278, 119\\

Held, Enrico V., \& Mould, Jeremy R. 1994 AJ 107, 1307\\

Impey, C., \& Bothun, G. 1997 ARA\&A 35 267\\

Impey, C., \& Bothun, G. 1989 ApJ 341, 89\\

Impey, C., Bothun, G., \& Malin, D.  1988, ApJ, 330, 634 (IBM)\\

Jerjen, H., Freeman, K.C., \& Binggeli 1998 AJ 116, 2873\\

Jerjen, H., \& Dressler, A 1997 A\&AS 124, 1\\

Kepner, J.V., Babul, A., \& Spergel, D.N. 1997 ApJ 487, 61\\

Knezek, P., M., Sembach, K. R., \& Gallagher, J. S.  1997 AAS 191, 8108\\

Kormendy, J.  1985 ApJ 295, 73\\

Kormendy, J.  1987 {\it Proceedings of the Eighth Santa Cruz Summer Workshop in
Astronomy and Astrophysics, Santa Cruz, CA, July 21-Aug. 1, 1986}
New York: Springer-Verlag\\

Lee, Myung G., Freedman, Wendy L., \& Madore, Barry F. 1993 AJ 106, 964L\\

McGaugh, S., Schombert, J., \& Bothun, G. 1995 AJ 109 2019\\

Meurer, G. R., Mackie, G., \& Carignan, C.  1994 AJ 107, 2021\\

Meylan, Georges, \& Prugniel, Philippe 1994; {\it ESO Conference and Workshop Proceedings}
Meylan,Georges \&  Prugniel, Philippe ed.,  Garching: European Southern Observatory\\

O'Neil, K.  1997, Ph.d. dissertation, University of Oregon, Eugene\\

O'Neil, K. Bothun, G. \& Impey C. 1998, AJ, 117\\

O'Neil, K. Bothun, G. \& Impey C. 1999, submitted to ApJS\\

Peterson, Ruth C. \& Caldwell, Nelson 1993, AJ, 105, 1411\\

Reed, B.C., Hesser, J., \& Shawl, S. 1988 PASP 100 545\\

Sage, L. J., Salzer, J. J., Loose, H.-H 1992 A\&A 265, 19\\

Sandage,A. \& Binggeli, B. 1984 AJ 89 919\\

Secker, J., Harris, W. E., \&  Plummer, J. D.  1997 PASP 109, 1377\\

Secker, Jeff, \& Harris, William E.  1996 ApJ 469, 623\\

Silk, Joseph, Wyse, Rosemary F. G., \& Shields, G. A.  1987 ApJ 322L, 59\\

Silva 1992, Ph.D. dissertation, University of Michigan, Ann Arbor\\

Spaans, Marco, \& Norman, Colin A.  1997 ApJ 488, 2\\

Sung, E-C, \etal 1998 ApJ 505 199\\

Tonry, John, \etal 1997 ApJ 475, 399\\

Tonry, John 1991 ApJ 373L 1\\

Tonry, John, \& Schneider, Donald P.  1988 AJ 96, 807\\

Ulmer,  M. P., Bernstein, G. M., Martin, D. R., Nichol, R. C., Pendleton, J. L.,
\& Tyson, J. A.  1996 AJ 112, 2517\\

Vader, J. Patricia 1987 ApJ 317, 128\\

Vader, J. Patricia, \& Chaboyer, Brian 1994 AJ 108, 1209\\

Whitemore, B. 1998, preprint\\

Worthey, G. 1994 ApJS 95, 107\\

Young, C. \& Currie, M.  1995 MNRAS 273 1141\\

\clearpage
\centerline{\bf Figures}

Figure~\ref{fig:wfpc}.  HST WFPC2 mosaicked images of V1L4 (a), V2L8 (b), and V7L3 (c) taken through the 
F814W (I band) filter with a 2200s exposure time.  These images are 2.6 arcminutes across.

Figure~\ref{fig:core}.  The nuclear regions of the three galaxies, V1L4, V2L8, and V7L3, respectively.
These images are each 49.8" across.

Figure~\ref{fig:sbrcore}.  Surface brightness profiles of the inner regions of the three Virgo 
galaxies.  Figure~\ref{fig:sbrcore}(a) shows V1L4, (b) shows the inner regions of V2L8 with the 
bright nucleus not included, and (c) shows the profile of V7L3.

Figure~\ref{fig:lum}.  Greyscale images of the central regions of the three galaxies in this 
study.  Figure~\ref{fig:lum}(a) shows V1L4, (b) and (c) show V2L8, and (d) shows 
V7L3.  White circles demarcate the inner and outer edges of the constant surface
brightness regions for each galaxy, 
while black contour line encircle the regions 1 $\sigma$ above the sky level.  All the 
images are 20'' across, except for Figure~\ref{fig:lum}(b), which shows the core of V2L8.
To allow for comparison between images, as section of V2L8 (shown by a black box in
Figure~\ref{fig:lum}(b)), which is 20'' across is shown in Figure~\ref{fig:lum}(c).
Note that these figures show the mosaicked images and are being shown for demonstration of
the studied areas only.  Mosaicked images were not used for the data analysis.

\clearpage
\centerline{\bf Tables}

Table 1.  The photometric and structural properties of the three Virgo galaxies as 
determined from ground based images. 

Table 2.  Luminosity fluctuations from the inner regions of the three galaxies.

Table 3. Stellar types used in the models 

Table 4:  Best fitting stellar populations to the observed pixel colors
and luminosity fluctuations.

\clearpage

\begin{deluxetable}{cccccccccc}
\tablewidth{0pt}
\tablehead{
\colhead{Galaxy}& \colhead{Names}& \colhead{RA}& \colhead{Dec}& \colhead{\Bmo}&
\colhead{\alp (")}& \colhead{B$_{27}$}& \colhead{D$_{27}$}& \colhead{B$-$V}& \colhead{V$-$I}\\
\colhead{(1)}& \colhead{(2)}& \colhead{(3)}& \colhead{(4)}& \colhead{(5)}& \colhead{(6)}&
\colhead{(7)}& \colhead{(8)}& \colhead{(9)}& \colhead{(10)}
}
\startdata
V1L4& VCC1582& 12:28:58.92& 12:54:29.7& 24.2\dag& \dag& 16.68& 74& 0.52& 0.71\nl
V2L8&        & 12:34:42.30& 14:13:22.4& 25.78&20.1& 18.33& 51& 0.46& 1.10\nl
V7L3& VCC1149& 12:28:58.98& 12:54:28.3& 25.07&19.1& 17.72& 60& 0.38\ddag& 0.79\ddag\nl

\multicolumn{10}{l}{\dag No line was fit to the exponential profile of the ground-based image,
and the central surface }\\
\multicolumn{10}{l}{\hskip 0.2in brightness value is approximate.}\\
\multicolumn{10}{l}{\ddag These colors were found at D=37''.}\\
\enddata
\end{deluxetable}

\clearpage
\begin{figure}[p]
%\centerline{
%\epsfxsize = 7.0in
%\epsffile{v1l4crm.ps}}
\caption{\label{fig:wfpc}a}
\end{figure}

%\clearpage
%\addtocounter{figure}{-1}
%\begin{figure}[p]
%\centerline{
%\epsfxsize = 7.0in
%\epsffile{v2l8crm.ps}}
%\caption{b}
%\end{figure}

%\clearpage
%\addtocounter{figure}{-1}
%\begin{figure}[p]
%\centerline{
%\epsfxsize = 7.0in
%\epsffile{v7l3crm.ps}}
%\caption{c}
%\end{figure}

%\clearpage
\begin{figure}[p]
%\centerline{
%\epsfxsize = 7.0in
%\epsffile{v1l4.ps}}
\caption{\label{fig:core}a}
\end{figure}

%\clearpage
%\addtocounter{figure}{-1}
%\begin{figure}[p]
%\centerline{
%\epsfxsize = 7.0in
%\epsffile{v2l8.ps}}
%\caption{b}
%\end{figure}

%\clearpage
%\addtocounter{figure}{-1}
%\begin{figure}[p]
%\centerline{
%\epsfxsize = 6.0in
%\epsffile{v7l3.ps}}
%\caption{c}
%\end{figure}

%\clearpage
\begin{figure}[p]
%\centerline{
%\epsfxsize = 5.0in
%\epsffile{iv1l4sb.ps}}
\caption{\label{fig:sbrcore}a}
\end{figure}

%\clearpage
%\addtocounter{figure}{-1}
%\begin{figure}[p]
%\centerline{
%\epsfxsize = 6.0in
%\epsffile{iv2l8sb.ps}}
%\caption{b}
%\end{figure}

%\clearpage
%\addtocounter{figure}{-1}
%\begin{figure}[p]
%\centerline{
%\epsfxsize = 6.0in
%\epsffile{iv7l3sb.ps}}
%\caption{c}
%\end{figure}

%\clearpage
\begin{figure}[p]
%\centerline{
%\epsfxsize = 5.0in
%\epsffile{iV1L4core2.ps}}
\caption{\label{fig:lum}a}
\end{figure}

%\clearpage
%\addtocounter{figure}{-1}
%\begin{figure}[p]
%\centerline{
%\epsfxsize = 6.0in
%\epsffile{iV2L8core.ps}}
%\caption{b}
%\end{figure}

%\clearpage
%\addtocounter{figure}{-1}
%\begin{figure}[p]
%\centerline{
%\epsfxsize = 6.0in
%\epsffile{iV2L8core2.ps}}
%\caption{c}
%\end{figure}

%\clearpage
%\addtocounter{figure}{-1}
%\begin{figure}[p]
%\centerline{
%\epsfxsize = 6.0in
%\epsffile{iV7L3core.ps}}
%\caption{d}
%\end{figure}

\end{document}